\documentclass[12pt]{article}
\usepackage{amsfonts,amsmath,amssymb,hyperref,graphicx,color}
\usepackage{physics}

\makeatletter
\def\amsbb{\use@mathgroup \M@U \symAMSb}
\makeatother
\usepackage{bbold}
\newcommand{\identity}{\mathbb{1}}

\newcommand{\be}{\begin{equation}}
\newcommand{\ee}{\end{equation}}
\newcommand{\bea}{\begin{eqnarray}}
\newcommand{\eea}{\end{eqnarray}}
\newcommand{\beas}{\begin{eqnarray*}}
\newcommand{\eeas}{\end{eqnarray*}}

\newcommand{\EtildeAB}{{\widetilde{E}_{AB}}}
\newcommand{\EtildeAC}{{\widetilde{E}_{AC}}}
\newcommand{\EtildeBC}{{\widetilde{E}_{BC}}}
\newcommand{\EtildeABC}{{\widetilde{E}_{ABC}}}

\begin{document}
\begin{titlepage}

\begin{center}

{\Large Entanglement groups for mixed states}

\vspace{12mm}

\renewcommand\thefootnote{\mbox{$\fnsymbol{footnote}$}}
Xiaole Jiang${}^{1,2}$\footnote{xjiang2@gradcenter.cuny.edu},
Daniel Kabat${}^{1,2}$\footnote{daniel.kabat@lehman.cuny.edu},
Gilad Lifschytz${}^{3}$\footnote{giladl@research.haifa.ac.il},
Aakash Marthandan${}^{1,2}$\footnote{amarthandan@gradcenter.cuny.edu}

\vspace{6mm}

${}^1${\small \sl Department of Physics and Astronomy} \\
{\small \sl Lehman College, City University of New York} \\
{\small \sl 250 Bedford Park Blvd.\ W, Bronx NY 10468, USA}

\vspace{4mm}

${}^2${\small \sl Graduate School and University Center, City University of New York} \\
{\small \sl  365 Fifth Avenue, New York NY 10016, USA}

\vspace{4mm}

${}^3${\small \sl Department of Physics and} \\
{\small \sl Haifa Research Center for Theoretical Physics and Astrophysics} \\
{\small \sl University of Haifa, Haifa 3498838, Israel}

\end{center}

\vspace{12mm}

\noindent
We extend an operational characterization of entanglement in terms of stabilizer groups from pure states
to mixed states.  For a density matrix $\rho_{AB}$, a stabilizer is a factorized unitary matrix $u_A \otimes u_B$
that, under conjugation, leaves $\rho_{AB}$ invariant.  The entanglement group is a quotient of the stabilizer group, in which one-party stabilizers
are considered trivial.  This definition relates the entanglement of a density matrix to the entanglement of its purification.
We give general properties of entanglement groups for mixed states, then discuss special properties for separable states.  For a separable state, the entanglement group may be non-trivial.  However it can only arise from
multi-party entanglement with the purifying system.

\end{titlepage}

\setcounter{footnote}{0}
\renewcommand\thefootnote{\mbox{\arabic{footnote}}}

\hrule
\tableofcontents
\bigskip
\hrule

\addtolength{\parskip}{8pt}

\section{Introduction\label{sect:introduction}}
Entanglement is one of the most fundamental properties of quantum mechanics.  Despite this, it remains rather mysterious.  At a practical level, although many entanglement
measures have been developed \cite{Horodecki:2009zz},  it is not straightforward to classify or quantify the entanglement of a given state.  At a conceptual level, it is hard to say what
entanglement is, exactly.  Certainly a tensor product state is not entangled, and all other states are entangled.  But what is the operational significance of having a state
that is not a tensor product?

In \cite{Jiang:2023uyj} we proposed an operational definition of entanglement for pure states, building on earlier works
\cite{Linden_1998,Carteret_1999,Carteret_2000,Vollbrecht_2001,Zurek1,Zurek2,Walck_2007,Tzitrin:2019fwe,Bernards:2022qhd}.  The definition led to a classification scheme for
entanglement in terms of groups, which we briefly recall.  Consider, for concreteness, a tripartite system with Hilbert space ${\cal H} = {\cal H}_A \otimes {\cal H}_B \otimes {\cal H}_C$ and
a pure state $\vert \psi \rangle \in {\cal H}$.  The group of local unitary transformations $U(d_A) \times U(d_B) \times U(d_C)$ acts on ${\cal H}$ as a separate unitary transformation
on each tensor factor.  We look for local unitary stabilizers, that is, elements of the local unitary group with the property that they leave the state invariant up to a phase.
\be
\label{LUstabilizer}
\big(u_A \otimes u_B \otimes u_C\big) \vert \psi \rangle = e^{i \theta} \vert \psi \rangle
\ee
These matrices form a group that we denote $S_{ABC}$.  This has a number of (normal) subgroups, consisting of matrices that act on fewer factors.  For example, there is a group of two-party stabilizers that act on $AB$,
\be
S_{AB} = \left\lbrace u_A \otimes u_B \otimes \identity_C \, \in S_{ABC} \right\rbrace
\ee
as well as a group of one-party stabilizers that just act on $A$.
\be
S_A = \left\lbrace u_A \otimes \identity_B \otimes \identity_C \, \in S_{ABC} \right\rbrace
\ee

These groups have clear operational meanings.  For example, $S_{AB}$ captures unitary transformations that can be performed on system $A$, which can be undone by
or are equivalent to unitary transformations on system $B$, without involving system $C$ in any way.  Intuitively, this should only be possible if $A$ and $B$ are entangled, however some of the transformations in $S_{AB}$
do not reflect entanglement.  In particular $S_A$ leaves the state invariant, without any need for a compensating transformation on system $B$, and similarly for $S_B$.  We are therefore led to
form the quotient group
\be
E_{AB} = S_{AB} / \left(S_A \times S_B\right)
\ee
We take this group to characterize two-party entanglement between systems $A$ and $B$, while leaving system $C$ aside.  We can likewise form two-party entanglement groups
$E_{AC}$ and $E_{BC}$.  We take non-trivial elements of these groups, i.e.\ elements other than the identity, to be a reflection of two-party entanglement between the corresponding systems.

What about the possibility of genuine three-party entanglement between systems $A$, $B$ and $C$?  This should be captured by $S_{ABC}$, however we wish to exclude transformations that are built up from a product of two-party stabilizers.  Thus we are led to define the
three-party entanglement group\footnote{The denominator is a group product, not a direct product.  This is well-defined because the group product of normal subgroups is a normal subgroup.}
\be
E_{ABC} = S_{ABC} / \left(S_{AB} \cdot S_{AC} \cdot S_{BC}\right)
\ee
We take non-trivial elements of this group, i.e.\ elements other than the identity, to be a reflection of three-party entanglement between systems $A$, $B$, $C$.

Note that these groups characterize the entanglement of the state $\vert \psi \rangle$ relative to the partition into three subsystems $A$, $B$, $C$.  It's important to recognize that this
does not capture all of the entanglement that may be present in $\vert \psi \rangle$.  In particular there may be unitary transformations on $A$ that can be undone by a unitary transformation
on the combined $BC$ system, however $u_{BC}$ does not factor into $u_B \otimes u_C$.  Denoting the combined $BC$ system by $(BC)$, such entanglement would be captured by the
group
\be
E_{A(BC)} = S_{A(BC)} / \left(S_A \times S_{(BC)}\right)
\ee

It's important to recognize that this group-based definition of entanglement, which we will refer to as $g$-entanglement if there is any possibility of confusion, is distinct from the usual notion
of entanglement which, as needed to avoid confusion, we will refer to as $u$-entanglement.  For example, a generic state for three parties cannot be tensor-factored, which means a generic
state has tri-partite $u$-entanglement.  In the $g$-entanglement scheme, this corresponds to the fact that a generic state has non-trivial entanglement groups $E_{A(BC)}$, $E_{B(AC)}$,
$E_{C(AB)}$.  However in the $g$-entanglement scheme, we do not regard this as true three-party entanglement.  That is captured by $E_{ABC}$, which is a much more restrictive definition.

To emphasize the physical meaning of a stabilizer such as (\ref{LUstabilizer}), consider writing the local unitary transformation in terms of a set of Hermitian generators.
\be
\label{LUstabilizer2}
\big(e^{i H_A} \otimes e^{i H_B} \otimes e^{i H_C}\big) \vert \psi \rangle = e^{i \theta} \vert \psi \rangle
\ee
If measurements of the observables $H_A$, $H_B$, $H_C$ are performed on the three separate parts of the system, the outcomes $h_A$, $h_B$, $h_C$ will be correlated
according to $h_A + h_B + h_C \equiv \theta \, {\rm mod} \, 2 \pi$.  Three-party entanglement, from this perspective, is responsible for correlations between $h_A$, $h_B$, $h_C$
that cannot be understood as a consequence of correlations involving fewer than three parties.

The goal of the present paper is to extend this scheme for characterizing and classifying entanglement to mixed states.
The entanglement group for a density matrix $\rho_{AB}$, which we denote $\EtildeAB$, is defined in section \ref{sect:definitions}.  One might expect that one could simply use
the pure-state entanglement groups associated with a purification of $\rho_{AB}$.  However the correct definition, given in (\ref{def1}), (\ref{def2}), (\ref{def3}) is slightly different:
it involves different quotients, since the purifying system is unobservable and must be
traced over.  In section \ref{sect:properties} we establish general properties of the entanglement groups for a density matrix, building
on results obtained in \cite{Jiang:2023uyj} for pure states.  In section \ref{sect:separable} we consider the entanglement
groups associated with a separable density matrix.  We will see that $\EtildeAB$ is not necessarily
trivial, due to the possibility of multi-party entanglement with the purifying system.

\section{Basic definitions\label{sect:definitions}}
To illustrate, we begin with a two-party density matrix $\rho_{AB}$.  The basic idea is to purify $\rho_{AB}$ to a state $\vert \psi \rangle_{ABC}$ and to introduce an appropriate notion of an entanglement
group associated with the purification.  After doing this, and showing that the entanglement groups are well-defined (independent of the purification), we return to relate the entanglement groups
directly to properties of $\rho_{AB}$.  The generalization to multipartite density matrices is straightforward and will be mentioned in passing.

Consider a bipartite Hilbert space ${\cal H}_A \otimes {\cal H}_B$.  An ensemble of states
\be
\lbrace \vert \ell \rangle_{AB} \in {\cal H}_A \otimes {\cal H}_B \rbrace \qquad \ell = 1,\ldots,L
\ee
with probabilities $p_\ell$ corresponds to a density matrix
\be
\rho_{AB} = \sum_{\ell = 1}^L p_\ell \, \vert \ell \rangle_{AB} \, {}_{AB} \langle \ell \vert
\ee
We're assuming the states $\vert \ell \rangle_{AB}$ are normalized, but they need not be orthogonal or complete in ${\cal H}_A \otimes {\cal H}_B$.

The density matrix can be purified by introducing an auxiliary Hilbert space ${\cal H}_C$ with an orthonormal basis $\lbrace \vert \ell \rangle_C \rbrace$ and
considering the state
\be
\vert \psi \rangle = \sum_{\ell = 1}^L \sqrt{p_\ell} \, \vert \ell \rangle_{AB} \otimes \vert \ell \rangle_C
\ee
Since ${}_C \langle \ell \vert \ell' \rangle_C = \delta_{\ell \ell'}$ we have ${\rm Tr}_C \left( \vert \psi \rangle \langle \psi \vert \right) = \rho_{AB}$.

Given the purification $\vert \psi \rangle_{ABC}$, we could define a variety of entanglement groups, as quotients of the various stabilizers $S_{ABC}$, $S_{AB}$, $S_{AC}$, $S_{BC}$,
$S_A$, $S_B$, $S_C$.  However we must ask which quotients have operational significance in terms of the original density matrix.  In this regard, remember that we do not
observe system $C$ and will eventually trace over it.  Therefore, the appropriate notion of a stabilizer is an element $\widetilde{s}_{AB} \in U(d_A) \times U(d_B)$ with the
property that it leaves the purification invariant up to a unitary on system $C$.
\be
\big(\widetilde{s}_{AB} \otimes \identity_C\big) \vert \psi \rangle = \big(\identity_{AB} \otimes u_C\big) \vert \psi \rangle
\ee
When tracing over system $C$, note that $u_C$ drops out, so such a stabilizer will leave the density matrix invariant.  In this way a stabilizer for the density matrix $\widetilde{s}_{AB}$ corresponds
to a stabilizer for the purification $\widetilde{s}_{AB} \otimes u_C^{-1}$.

We can obtain the stabilizer groups for $\rho_{AB}$, denoted with a tilde, by acting with $AB$ projection operators on the stabilizer groups for the purification.\footnote{The projection is defined by
$\pi_{AB}(u_A,u_B,u_C) = (u_A,u_B)$.}
\bea
\nonumber
&& \widetilde{S}_{AB} = \pi_{AB}\big(S_{ABC}\big) \\
&& \widetilde{S}_A = \pi_{AB}\big(S_{AC}\big) \\
\nonumber
&& \widetilde{S}_B = \pi_{AB}\big(S_{BC}\big)
\eea
Given these stabilizer groups, we can construct an entanglement group for the density matrix
\be
\label{EtildeAB}
\EtildeAB = \widetilde{S}_{AB} / \big(\widetilde{S}_A \times \widetilde{S}_B\big) = \pi_{AB}\big(S_{ABC} / \left(S_{AC} \cdot S_{BC}\right)\big)
\ee
This is the entanglement group that we associate with a two-party density matrix $\rho_{AB}$.  To make a few comments:
\begin{itemize}
\item
From the point of view of the purification, $\widetilde{S}_{AB}$ allows a unitary transformation on system $C$.  This means it includes both
the  two-party $AB$ entanglement and the three-party $ABC$ entanglement that is present in the purification.  If we had access to system $C$ we could distinguish these possibilities
by examining the entanglement groups associated with the pure state $\vert \psi \rangle$, namely
\bea
\nonumber
&& E_{AB} = S_{AB} / \left(S_A \times S_B\right) \\[5pt]
\label{distinguish}
&& E_{ABC} = S_{ABC} / \left(S_{AB} \cdot S_{AC} \cdot S_{BC}\right)
\eea
However, the assumption is that we do not have access to system $C$, so if we are just given the density matrix, we cannot make this distinction in a meaningful way.  Instead, the only
physically meaningful entanglement group associated with $\rho_{AB}$ is $\EtildeAB$.
\item
As mentioned above, the groups $E_{AB}$ and $E_{ABC}$ do not capture all of the
entanglement involving $A$ and $B$ which may be present in the purification.  There could also be stabilizers $S_{A(BC)}$ that act with a unitary on the combined $(BC)$ system.  We will return to this point in section \ref{sect:A(BC)}.
\item
The generalization to multipartite mixed states is straightforward.  For example, consider a tripartite density matrix $\rho_{ABC}$.  This has a purification $\vert \psi \rangle_{ABCD}$
and we can define entanglement groups
\bea
\nonumber
&&\EtildeAB = \pi_{ABC}\big(S_{ABD} / \left(S_{AD} \cdot S_{BD}\right)\big) \\
\nonumber
&&\EtildeAC = \pi_{ABC}\big(S_{ACD} / \left(S_{AD} \cdot S_{CD}\right)\big) \\
\label{EtildeABC}
&&\EtildeBC = \pi_{ABC}\big(S_{BCD} / \left(S_{BD} \cdot S_{CD}\right)\big) \\
\nonumber
&&\EtildeABC = \pi_{ABC}\big(S_{ABCD} / \left(S_{ABD} \cdot S_{ACD} \cdot S_{BCD}\right)\big)
\eea
\end{itemize}

Since the definitions (\ref{EtildeAB}), (\ref{EtildeABC}) are in terms of a purification, we must show that these entanglement groups are
well-defined, that is, that they are independent of the choice of purification.  Every density
matrix has a minimal purification, constructed as follows.  Start by diagonalizing the density matrix.
\be
\rho_{AB} = \sum_{i = 1}^r p_i \, \vert i \rangle_{AB} {}_{AB} \langle i \vert
\ee
Here $r$ is the rank of $\rho_{AB}$.  A minimal purification is then
\be
\vert \psi \rangle = \sum_{i = 1}^r \sqrt{p_i} \, \vert i \rangle_{AB} \otimes \vert i \rangle_C
\ee
Any other purification of $\rho_{AB}$ is related to this one by \cite{nielsen2010quantum}
\begin{enumerate}
\item
Embedding the orthonormal vectors $\vert i \rangle_C$ into a (sufficiently large) Hilbert space ${\cal H}_{C'}$;
\item
Performing a unitary transformation on ${\cal H}_{C'}$.
\end{enumerate}
For later reference, two different ensembles $\lbrace (p_\ell,\, \vert \ell \rangle_{AB}) \rbrace$ and $\lbrace (q_m,\, \vert m \rangle_{AB}) \rbrace$ can generate
exactly the same density matrix \cite{nielsen2010quantum}.  The ensembles lead to nominally different purifications,
\bea
\nonumber
&& \vert \psi \rangle = \sum_{\ell = 1}^L \sqrt{p_\ell} \, \vert \ell \rangle_{AB} \otimes \vert \ell \rangle_C \\
\label{freedom}
&& \vert \psi' \rangle = \sum_{m = 1}^M \sqrt{q_m} \, \vert m \rangle_{AB} \otimes \vert m \rangle_{C'}
\eea
which are however related in the manner described (if necessary, by embedding the vectors in the larger of ${\cal H}_C$ and ${\cal H}_{C'}$, then performing
a unitary transformation on the auxiliary Hilbert space).

Are the entanglement groups associated with a density matrix well-defined?  For $\rho_{AB}$, entanglement groups are defined in terms of stabilizer groups such as
$S_{AC}$ and $S_{ABC}$.  These groups transform in a simple way under unitary transformations on system $C$.\footnote{Likewise $S_{A(BC)}$ behaves
nicely under unitary transformations on the combined $(BC)$ system, which includes unitary transformations on system $C$ as a subgroup.}
\bea
\nonumber
&& \vert \psi \rangle \rightarrow \left(\identity_A \otimes \identity_B \otimes u_C\right) \vert \psi \rangle \equiv {\cal U}_C \vert \psi \rangle \\
&& S_{AC} \rightarrow {\cal U}_C S_{AC} \, {\cal U}_C^\dagger \\
\nonumber
&& S_{ABC} \rightarrow {\cal U}_C S_{ABC} \, {\cal U}_C^\dagger
\eea
These transformations leave the abstract groups $S_{AC}$ and $S_{ABC}$ invariant, although they change the way they're embedded in
$U(d_A) \times U(d_B) \times U(d_C)$.  So the entanglement groups are well-defined, independent of the choice of purification.

At this stage we have a definition of entanglement groups for a density matrix $\rho_{AB}$ in terms of stabilizers of the purification $\vert \psi \rangle$.  What do these stabilizers
mean in terms of the density matrix itself?  To illustrate, consider a stabilizer
\be
s_{ABC} = u_A \otimes u_B \otimes u_C \in S_{ABC}
\ee
which by definition means
\be
s_{ABC} \vert \psi \rangle = e^{i \theta} \vert \psi \rangle
\ee
or equivalently
\be
\big(u_A \otimes u_B \otimes \identity_C\big) \vert \psi \rangle = e^{i \theta} \big(\identity_{AB} \otimes u_C^{-1}\big) \vert \psi \rangle
\ee
Tracing out system $C$, and denoting $\widetilde{s}_{AB} = u_A \otimes u_B$, we have
\be
\label{RhoStabilizer}
\widetilde{s}_{AB} \, \rho_{AB} \, \widetilde{s}_{AB}^{\,\,\dagger} = \rho_{AB}
\ee
So the density matrix is invariant under conjugation by $\widetilde{s}_{AB}$.

Conversely, given a stabilizer $\widetilde{s}_{AB}$ satisfying (\ref{RhoStabilizer}), it can be lifted to an element of $S_{ABC}$ for the purification.  To see this, suppose
$\lbrace (p_\ell,\, \vert \ell \rangle_{AB}) \rbrace$ is an ensemble of states that correspond to the density matrix $\rho_{AB}$.  Given a stabilizer (\ref{RhoStabilizer}), it
follows that $\lbrace (p_\ell,\, \widetilde{s}_{AB} \vert \ell \rangle_{AB}) \rbrace$ corresponds to exactly the same density matrix.  Then, due to the uniqueness discussed
around (\ref{freedom}), the two purifications are related by a unitary transformation $u_C$ on the auxiliary system.  That is, there is a stabilizer for the purification
$\widetilde{s}_{AB} \otimes u_C \in S_{ABC}$.

This means we can define the group $\widetilde{S}_{AB}$ as the subgroup of $U(d_A) \times U(d_B)$ with the property that the density matrix is invariant under conjugation
by every element of $\widetilde{S}_{AB}$.
\be
\label{def1}
\widetilde{s}_{AB} \, \rho_{AB} \, \widetilde{s}_{AB}^{\,\,\dagger} = \rho_{AB} \qquad \hbox{\rm for all $\widetilde{s}_{AB} \in \widetilde{S}_{AB}$}
\ee
This has a normal subgroups $\widetilde{S}_A$, $\widetilde{S}_B$ which act trivially on one of the subsystems.
\bea
\nonumber
&& \widetilde{S}_A = \left\lbrace \hbox{$u_A \otimes \identity_B \in \widetilde{S}_{AB}$ such that $\big(u_A \otimes \identity_B\big) \rho_{AB} \big(u_A \otimes \identity_B\big)^\dagger = \rho_{AB}$}\right\rbrace \\
\label{def2}
&& \widetilde{S}_B = \left\lbrace \hbox{$\identity_A \otimes u_B \in \widetilde{S}_{AB}$ such that $\big(\identity_A \otimes u_B\big) \rho_{AB} \big(\identity_A \otimes u_B\big)^\dagger = \rho_{AB}$}\right\rbrace
\eea
This lets us construct the entanglement group
\be
\label{def3}
\EtildeAB = \widetilde{S}_{AB} / (\widetilde{S}_A \times \widetilde{S}_B)
\ee
directly from the density matrix, without reference to a purification.  Note that this construction is operationally meaningful, in the sense that it involves a physical operation (unitary conjugation) on the density matrix.

\section{Properties of entanglement groups\label{sect:properties}}
Having defined entanglement groups for density matrices, we'd like to explore some of their properties.  In what follows we're mostly concerned
with the action of the entanglement groups on individual systems.  For this reason we introduce projection operators $\pi_A, \, \pi_B, \, \pi_C, \, \ldots$ that project
onto systems $A, \, B, \, C, \, \ldots$.  For example, given $g = (g_A,g_B) \in \EtildeAB$, we have $\pi_A(g) = g_A$.

There are two main questions we'd like to address.
\begin{enumerate}
\item
Consider a two-party density matrix $\rho_{AB}$.  Let $g_A \in \pi_A(\EtildeAB)$.  Is there a unique element $g_{AB} \in \EtildeAB$ that projects
to give $g_A$?  If so, how are the actions of $g_{AB}$ on systems $A$ and $B$ related?
\item
Consider a three-party density matrix $\rho_{ABC}$.  Let $g_A \in \pi_A(\EtildeAB)$.  Could $g_A$ appear in the projection of any other entanglement groups?
That is, could $g_A$ also be an element of $\pi_A(\EtildeAC)$ or of $\pi_A(\EtildeABC)$?
\end{enumerate}
In what follows we show that:
\begin{enumerate}
\item
There is a unique $g_{AB} \in \EtildeAB$ that projects to any given $g_A \in \pi_A(\EtildeAB)$.  It acts isomorphically on systems $A$ and $B$.
\item
A transformation $g_A$ can appear in both $\pi_A(\EtildeAB)$ and $\pi_A(\EtildeAC)$, but only if it is an element of the center of both groups.
It cannot appear in the projection of both a two-party entanglement group (meaning $\EtildeAB$, $\EtildeAC$) and a three-party entanglement group $\EtildeABC$.
\end{enumerate}
The discussion here closely follows the treatment of pure states in \cite{Jiang:2023uyj}.

\subsection{Isomorphism theorems\label{sect:isomorphism}}
For a two-party density matrix $\rho_{AB}$, we wish to show that there is a unique $g_{AB} \in \EtildeAB$ that projects to any given $g_A \in \pi_A(\EtildeAB)$, and
that $g_{AB}$ acts isomorphically on systems $A$ and $B$.  We begin with some suggestive observations at the level of stabilizers, then turn to entanglement groups.

Purify $\rho_{AB}$ to a state $\vert \psi \rangle_{ABC}$, and consider two stabilizers which happen to have the same action on system $A$.
\bea
\nonumber
&& s_1 = (s_A,s_B,s_C) \in S_{ABC} \\[5pt]
&& s_2 = (s_A,s_B',s_C') \in S_{ABC}
\eea
It follows that
\be
s_1 s_2^{-1} = (\identity_A, s_B s_B'^{\,-1}, s_C s_C'^{\,-1})
\ee
is a stabilizer which only acts on systems $B$ and $C$, so $s_1 s_2^{-1} \in S_{BC}$.  When we quotient to form $\EtildeAB$, elements of $S_{BC}$ are mapped to the identity.
So $s_1$ and $s_2$ represent the same equivalence class in $\EtildeAB$.  This means $s_A$ determines a unique element $g_{AB} \in \EtildeAB$, and hence a unique
element $g_B \in \pi_B(\EtildeAB)$.

This observation about stabilizers is suggestive, but does not quite establish the desired result.  We'd like to make a statement, not about $s_A$, but rather about an
equivalence class $[s_A] \in \pi_A(\EtildeAB)$.  For this we first introduce some convenient notation.  Let
\be
G = S_{ABC} \qquad\quad N = S_{AC} \cdot S_{BC}
\ee
with projections
\bea
\nonumber
G_A = \pi_A(G) \qquad\quad N_A = \pi_A(N) \\[5pt]
G_B = \pi_B(G) \qquad\quad N_B = \pi_B(N)
\eea
The entanglement group $\EtildeAB$ is constructed as
\be
\EtildeAB = \pi_{AB}(G / N)
\ee
We claim there are group isomorphisms
\bea
\nonumber
&& \theta_A \, : \, \EtildeAB \rightarrow G_A / N_A \\
&& \theta_B \, : \, \EtildeAB \rightarrow G_B / N_B
\eea
which make
\be
\label{isomorphism}
\EtildeAB \approx G_A / N_A \approx G_B / N_B
\ee
This means we can present $\EtildeAB$ in the form
\be
\EtildeAB = \left\lbrace \, \Big(\,\theta_A(t),\, \theta_B(t)\,\Big) \, \Big\vert \, t \in \EtildeAB \, \right\rbrace
\ee
So there is a unique element of $\EtildeAB$ that projects to any given $[s_A]$ or $[s_B]$, and it acts isomorphically on systems
$A$ and $B$.

The proof of (\ref{isomorphism}) is as follows.  The isomorphism $G_A / N_A \approx G_B / N_B$ is given as Theorem 1
in appendix B of \cite{Jiang:2023uyj}.  We also have an isomorphism\footnote{Proof: let $H = \left\lbrace \, \Big( \, [s_A],\,[s_B],\,[s_C] \, \Big) \, \in  {G_A \over N_A} \times {G_B \over N_B} \times {G_C \over N_C} \, \Big\vert \,
\big(s_A,s_B,s_C\big) \in G \, \right\rbrace$ and define a map $\phi \, : \, G \rightarrow H$ by $\phi\big((s_A,s_B,s_C)\big) = \big( \, [s_A],\,[s_B],\,[s_C] \, \big)$.  Then ${\rm ker} \, \phi = N$ and
an isomorphism theorem gives $G / {\rm ker} \phi \approx H$.}
\be
G / N \approx \left\lbrace \, \Big( \, [s_A],\,[s_B],\,[s_C] \, \Big) \, \in  {G_A \over N_A} \times {G_B \over N_B} \times {G_C \over N_C} \, \Big\vert \,
(s_A,s_B,s_C) \in G \, \right\rbrace
\ee
Acting on this with $\pi_{AB}$ we have the isomorphisms given in (\ref{isomorphism}).

\subsection{Restrictions on sharing\label{sect:sharing}}
Now we turn to a three-party density matrix $\rho_{ABC}$ and ask: can a given element $g_A \in \pi_A(\EtildeAB)$ appear in the
projection of more than one entanglement group?  Could it also appear in $\pi_A(\EtildeABC)$?  What about $\pi_A(\EtildeAC)$?

We begin by disposing of the first possibility: if $g_A$ appears in the projection of a two-party entanglement group, it cannot also
appear in the projection of a three-party group.  To see this, suppose there are stabilizers
\bea
\nonumber
&& s_1 = \big(s_A,s_B,\identity_C,s_D\big) \in S_{ABD} \\
&& s_2 = \big(s_A,s_B',s_C',s_D'\big) \in S_{ABCD}
\eea
Then $s_1^{-1} s_2 = \big(\identity_A, s_B^{-1}s_B', s_C', s_D^{-1} s_D' \big) \in S_{BCD}$, which means we can write $s_2$ as a product of an element of $S_{ABD}$ with an element of $S_{BCD}$.
\be
s_2 = \big(s_A,s_B,\identity_C,s_D\big) \cdot \big(\identity_A, s_B^{-1}s_B', s_C', s_D^{-1} s_D' \big)
\ee
Recall that $\EtildeABC$ involves a quotient by $S_{ABD}$ and $S_{BCD}$.  So written in this way, we see that $s_2$ maps to the identity element of $\EtildeABC$.

Could $g_A$ also appear in $\pi_A(\EtildeAC)$?  This is a bit more subtle.  Consider two stabilizers
\bea
\nonumber
&& s_1 = \big(s_A,s_B,\identity_C,s_D\big) \in S_{ABD} \\
&& s_2 = \big(s_A',\identity_B,s_C',s_D'\big) \in S_{ACD}
\eea
Their group commutator is
\be
s_1 s_2 s_1^{-1} s_2^{-1} = \big(s_A s_A' s_A^{-1} s_A'^{\,-1}, \identity_B, \identity_C, s_D s_D' s_D^{-1} s_D'^{\,-1}\big) \in S_{AD}
\ee
The commutator becomes trivial in both $\EtildeAB$ and $\EtildeAC$, since those groups involve a quotient by $S_{AD}$.
Projecting onto system $A$, we see that $\pi_A(\EtildeAB)$ and $\pi_A(\EtildeAC)$ commute element-by-element.  This means that a single
element $g_A$ could appear in both groups, but only if it is an element of the center of both.
\be
\label{Z}
g_A \in Z\big(\pi_A(\EtildeAB)\big) \cap Z\big(\pi_A(\EtildeAC)\big)
\ee
In this sense, $g$-entanglement is not strictly monogamous: an element of the common center (\ref{Z}) can be thought of as reflecting
either $AB$ entanglement or $AC$ entanglement.

\section{Separable states\label{sect:separable}}
When a density matrix $\rho_{AB}$ is separable, there are only classical correlations between measurements on systems $A$ and
$B$.  This is taken to mean there's no entanglement between $A$ and $B$ \cite{Werner:1989zz}.  Our goals here are to translate
separability into the language of stabilizer groups, and to understand what it means from this perspective.
We will see that, although separability places severe restrictions, it does not quite imply that $\EtildeAB$ is trivial,
because the purification could have three-party entanglement with the auxiliary system.

Consider a bipartite Hilbert space ${\cal H}_A \otimes {\cal H}_B$.  A separable density matrix $\rho_{AB}$ has the form
\be
\label{separable}
\rho_{AB} = \sum_\ell p_\ell \, \rho_A^{(\ell)} \otimes \rho_B^{(\ell)}
\ee
with $p_\ell > 0$ and $\sum_\ell p_\ell = 1$.  We can take $\rho_A^{(\ell)}$ and $\rho_B^{(\ell)}$ to be pure-state density matrices \cite{Horodecki:1997vt,Horodecki:2009zz}, so that
\be
\label{separable2}
\rho_{AB} = \sum_{\ell = 1}^L p_\ell \, \big(\vert \ell \rangle_A \, {}_A \langle \ell \vert \big) \otimes \big(\vert \ell \rangle_B \, {}_B \langle \ell \vert \big)
\ee
It follows from Carath\'odory's theorem that $L \leq d_{AB}^2 = (d_A d_B)^2$ \cite{Horodecki:1997vt,Horodecki:2009zz}, where the dimensions $d_A = {\rm dim} \, {\cal H}_A$,
$d_B = {\rm dim} \, {\cal H}_B$ are assumed finite.

In what follows we first discuss purification of $\rho_{AB}$, by introducing an auxiliary Hilbert space ${\cal H}_C$.  Then we examine
the stabilizers and entanglement groups associated with the purification.  We will consider both
\begin{itemize}
\item
the group $E_{AB}$, which captures entanglement between $A$ and $B$ while leaving out the auxiliary system;
\item
the group $E_{A(BC)}$, which captures entanglement between $A$ and the combined $(BC)$ system.
\end{itemize}
Along the way we will see that, although the entanglement group $\EtildeAB$ associated with the density matrix itself can be non-trivial,
it can only arise from three-party entanglement between $A$, $B$ and the purifying system $C$.

\subsection{Purifying a separable density matrix}
Given a separable density matrix $\rho_{AB}$, we can purify it to a pure state of a tripartite system $ABC$ by introducing
\be
\label{pure}
\vert \psi \rangle = \sum_\ell \sqrt{p_\ell} \, \vert \ell \rangle_A \otimes \vert \ell \rangle_B \otimes \vert \ell \rangle_C
\ee
Here $\lbrace \vert \ell \rangle_C \rbrace$ is a basis for the Hilbert space ${\cal H}_C$, which we take to be orthonormal.
\be
{}_C\langle \ell \vert \ell' \rangle_C = \delta_{\ell \ell'} \qquad \ell,\,\ell' = 1,\ldots,L
\ee
A purification of this form is only possible for a separable density matrix.  A general superposition of factorized states
\be
\vert \psi \rangle = \sum_\ell \alpha_\ell \, \vert \ell \rangle_A \otimes \vert \ell \rangle_B \otimes \vert \ell \rangle_C \qquad \alpha_\ell \in {\amsbb C}
\ee
leads to
\be
\rho_{AB} = \sum_\ell \vert \alpha_\ell \vert^2 \big(\vert \ell \rangle_A \, {}_A \langle \ell \vert \big) \otimes \big(\vert \ell \rangle_B \, {}_B \langle \ell \vert \big)
\ee
So separability --  including the fact that the coefficients $p_\ell$ in (\ref{separable}) are all positive -- is essential.

A few comments on this purification are in order.  First, note that there's no reason to expect the vectors $\vert \ell \rangle_A$ to be linearly independent in ${\cal H}_A$, nor is there any reason to expect them to
span all of ${\cal H}_A$.  Likewise for $\vert \ell \rangle_B$ and ${\cal H}_B$.  Next, note that the auxiliary Hilbert space ${\cal H}_C$ we introduced has dimension ${\rm dim} \, {\cal H}_C = L$.  The minimum dimension required to purify
a density matrix is equal to the rank of the density matrix.  Generically $L$ is bigger than the rank of $\rho_{AB}$, so generically we're using a non-minimal purification.
However, as discussed previously, it is related to any other purification by a unitary transformation on system $C$.

\subsection{Entanglement group for a separable density matrix}
Suppose a density matrix is separable.  What does separability imply for the associated entanglement group?  One might expect that
\be
\label{EtildeAB2}
\EtildeAB = \pi_{AB}\big(S_{ABC} / \left(S_{AC} \cdot S_{BC}\right)\big)
\ee
should be trivial.  It turns out this is not quite the case.

To understand what separability implies, we first purify $\rho_{AB}$ to a state of the form (\ref{pure}).  In appendix \ref{appendix:twoparty} we show that any two-party stabilizer of the purification $s_{AB}$ can be written as a product
$s_{AC} s_{BC}$.  Such stabilizers can give the pure state $\vert \psi \rangle$ a two-party entanglement group, since there is no reason for
\be
E_{AB} = S_{AB} / \left(S_A \times S_B\right)
\ee
to be trivial.  However such stabilizers do not contribute to the entanglement group for the density matrix $\EtildeAB$: due to the quotient
by $S_{AC} \cdot S_{BC}$, such stabilizers map to the identity.  Although two-party
entanglement in the purification does not contribute to $\EtildeAB$, it is still possible for $\EtildeAB$ to be non-trivial, because there could be
three-party $g$-entanglement in the purification.

To illustrate this, consider the generalized GHZ state
\be
\vert \psi \rangle = a \vert 000 \rangle + b \vert 111 \rangle
\ee
For generic values of $a$ and $b$ the entanglement groups are \cite{Jiang:2023uyj}\footnote{Here $\varnothing$ denotes the trivial group.  The generalized GHZ state has tri-partite $u$-entanglement,
which in the $g$-entanglement approach corresponds to having non-trivial groups $E_{A(BC)} = E_{B(AC)} = E_{C(AB)} = U(1)$.  However it does not have bi-partite $u$-entanglement
since the reduced density matrices are separable.  For a discussion of $g$-entanglement vs.\ $u$-entanglement in GHZ see \cite{Jiang:2023uyj}.}
\be
E_{AB} = U(1) \qquad\quad E_{ABC} = \varnothing
\ee
Tracing out system $C$ leads to a separable density matrix
\be
\rho_{AB} = \vert a \vert^2 \, \vert 00 \rangle \langle 00 \vert + \vert b \vert^2 \, \vert 11 \rangle \langle 11 \vert
\ee
with $\EtildeAB = \varnothing$.  The absence of three-party entanglement in the purification means that, as argued in general above,
$\EtildeAB$ is trivial.

By contrast, the standard GHZ state with $a = b = 1/\sqrt{2}$ has an additional three-party stabilizer.
\be
x = \begin{pmatrix}
0 & 1 \\
1 & 0
\end{pmatrix}
\otimes
\begin{pmatrix}
0 & 1 \\
1 & 0
\end{pmatrix}
\otimes
\begin{pmatrix}
0 & 1 \\
1 & 0
\end{pmatrix} \in S_{ABC}
\ee
This enhances $E_{ABC}$ from $\varnothing$ to ${\amsbb Z}_2$.  As a result, $\EtildeAB$ is also enhanced to ${\amsbb Z}_2$.  This illustrates the fact that three-party entanglement
with the purifying system can lead to a non-trivial $\EtildeAB$.

As a further comment, one could imagine starting with a pure state $\vert \psi \rangle_{ABC}$ and obtaining $\rho_{AB}$ as the reduced density matrix that describes measurements
performed on the $AB$ system.  In this case system $C$ is in principle observable, so as discussed around (\ref{distinguish}), more information is in principle available.  In particular, in addition to
the entanglement group $\EtildeAB$ that is associated with the density matrix, we could consider the two-party entanglement group $E_{AB}$ that is associated with the pure state
$\vert \psi \rangle_{ABC}$.
(Recall that $E_{AB}$ is the entanglement group built from stabilizers that do not act on system $C$.  Given the density matrix $\rho_{AB}$
this is not meaningful, but given the pure state $\vert \psi \rangle_{ABC}$ it is.)  $E_{AB}$ has a very particular structure.
We have seen that, if $\rho_{AB}$ is separable, all $AB$ stabilizers
are built up as a product of $AC$ and $BC$ stabilizers.  As shown in appendix \ref{appendix:twoparty}, this means that in the $g$-entanglement scheme $AB$ entanglement can be thought of as a combination of $AC$ entanglement with $BC$ entanglement.

\subsection{Structure of $A$ -- $(BC)$ entanglement in the purification\label{sect:A(BC)}}
Here we consider entanglement between system $A$ and the combined $(BC)$ system, captured by the group $E_{A(BC)}$.  What does separability have to say about such entanglement?

First, note that $\rho_{AB}$ is separable if and only if there is a controlled unitary\footnote{a unitary transformation ${\cal U}_{BC}$ that is block diagonal, as in the second
factor of (\ref{block}), controlled by the state of system $C$} that acts on the combined $(BC)$ system and
completely disentangles $B$.  To construct ${\cal U}_{BC}$, consider the purification of a separable density matrix to
\be
\label{pure2}
\vert \psi \rangle = \sum_{\ell = 1}^L \sqrt{p_\ell} \, \vert \ell \rangle_A \otimes \vert \ell \rangle_B \otimes \vert \ell \rangle_C
\ee
Choose any fixed vector $\vert \chi \rangle_B \in {\cal H}_B$ and choose a set of unitary matrices $u_B^{(\ell)}$ with the property that for each $\ell$
\be
u_B^{(\ell)} \vert \ell \rangle_B = \vert \chi \rangle_B
\ee
Since we don't specify how $u_B^{(\ell)}$ acts on any vector orthogonal to $\vert \ell \rangle_B$, there is quite a bit of freedom in choosing these matrices.\footnote{There is
a unitary transformation that maps any orthonormal basis to any other orthonormal basis.}  Then define
\bea
\nonumber
{\cal U}_{BC} & = & \sum_{\ell = 1}^L \identity_A \otimes u_B^{(\ell)} \otimes \vert \ell \rangle_C {}_C \langle \ell \vert \\
\label{block}
& = & \identity_A \otimes \left(\begin{array}{ccc} u_B^{(1)} & & \\ & \ddots & \\ & & u_B^{(\ell)} \end{array}\right)_{BC}
\eea
This is the controlled unitary we referred to previously.  It has the property that
\be
\label{factorized}
{\cal U}_{BC} \vert \psi \rangle = \sum_{\ell = 1}^L \sqrt{p_\ell} \, \vert \ell \rangle_A \otimes \vert \chi \rangle_B \otimes \vert \ell \rangle_C
\ee
A controlled unitary for the Werner state is given in appendix \ref{appendix:Werner}.

The key point is that the right hand side of (\ref{factorized}) is a tensor product.  With a slight abuse of notation we can write it as
\be
{\cal U}_{BC} \vert \psi \rangle = \vert \chi \rangle_B \otimes \Big(\sum_{\ell = 1}^L \sqrt{p_\ell} \, \vert \ell \rangle_A \otimes \vert \ell \rangle_C\Big)
\ee
The unitary ${\cal U}_{BC}$ shifts all of the entanglement of $A$ with the combined $BC$ system so that it becomes entanglement just between $A$ and $C$.  As shown in appendix \ref{appendix:UBC},
the entanglement groups for the state ${\cal U}_{BC} \vert \psi \rangle$ reflect the fact that system $B$ has been completely disentangled.

Note that we can go on to construct a controlled unitary ${\cal U}_{AC}$ which completely disentangles system $A$.  By combining the two transformations, we can map the original state to a tensor product.
\be
{\cal U}_{AC} \, {\cal U}_{BC} \vert \psi \rangle = \vert \chi \rangle_A \otimes \vert \chi \rangle_B \otimes \Big(\sum_{\ell = 1}^L \sqrt{p_\ell} \, \vert \ell \rangle_C \Big)
\label{sepprep}
\ee
This characterization of pure states whose reduced density matrices are separable is related to the usual notion of how one makes a separable
density matrix.  From (\ref{sepprep}) there is a controlled unitary ${\cal U}_{ABC}={\cal U}^{-1}_{AC}{\cal U}^{-1}_{BC} $ with party $C$ as the control,
for which
\be
{\cal U}_{ABC} \, \vert \, \hbox{\rm tensor product state} \, \rangle = \vert \psi \rangle
\label{dmp}
\ee
The usual  procedure for making a separable $\rho_{AB}$ is for party $C$ to generate a random number $\ell \in \lbrace 1,\ldots,L \rbrace$  with probability $p_\ell$, and to
communicate the result to parties $A$ and $B$ who then prepare their states accordingly.  Party $C$ could generate the random number by making
a measurement of $\ell$ in the state $\sum_{\ell = 1}^L \sqrt{p_\ell} \, \vert \ell \rangle_C$.  Due to the delayed measurement principle \cite{aharonov1998quantumcircuitsmixedstates,GUREVICH202221}, an alternate procedure is to apply a controlled unitary, then at the end measure
(or trace over) system $C$.  This results in identical statistical outcomes for
the $AB$ system. This alternate procedure can be implemented by applying ${\cal U}_{ABC}$ as in (\ref{dmp}) then measuring or tracing over $C$.

\bigskip
\goodbreak
\centerline{\bf Acknowledgements}
\noindent
We are grateful to Dorit Aharonov for many valuable discussions and suggestions.
The work of XJ, DK and AM was supported by U.S.\ National Science Foundation grant PHY-2112548.
GL was supported in part by the Israel Science Foundation under grant 447/17.

\appendix
\section{Two-party stabilizers for a separable state\label{appendix:twoparty}}
Here we consider the purification of a separable density matrix given in (\ref{pure}).
\be
\label{pure3}
\vert \psi \rangle = \sum_\ell \sqrt{p_\ell} \, \vert \ell \rangle_A \otimes \vert \ell \rangle_B \otimes \vert \ell \rangle_C
\ee
We want to understand what separability implies for a two-party stabilizer that acts on systems $A$ and $B$.

To this end, suppose there's a stabilizer $s_{AB} = u_A \otimes u_B \otimes \identity_C \in S_{AB}$.  This means
\be
\label{uAuB1}
(u_A \otimes u_B \otimes \identity_C) \vert \psi \rangle = e^{i \theta} \vert \psi \rangle \qquad \hbox{\rm for some phase $\theta$}
\ee
What can we say about the structure of the stabilizer, given the form of the state $\vert \psi \rangle$?  By acting on (\ref{uAuB1}) with ${}_C\langle \ell \vert$ we see that
\be
\label{uAuBonl}
u_A \vert \ell \rangle_A \otimes u_B \vert \ell \rangle_B = e^{i \theta} \vert \ell \rangle_A \otimes \vert \ell \rangle_B \qquad \hbox{\rm same phase $\theta$ for every $\ell$}
\ee
If a non-zero vector $x$ can be written as a tensor product in two different ways, say $x = y_1 \otimes y_2 = z_1 \otimes z_2$, it follows that the vectors are proportional,
$y_1 = c z_1$ and $y_2 = {1 \over c} z_2$.  Since $u_A$ and $u_B$ are unitary they can't change the magnitude of a vector, but they can introduce phases.  So from (\ref{uAuBonl}) we must have
\be
u_A \vert \ell \rangle_A = e^{i \alpha_\ell} \vert \ell \rangle_A \qquad u_B \vert \ell \rangle_B = e^{i \beta_\ell} \vert \ell \rangle_B \quad {\rm with} \quad \alpha_\ell + \beta_\ell = \theta
\ee
In other words $\vert \ell \rangle_A$, $\vert \ell \rangle_B$ are eigenvectors of $u_A$, $u_B$ with eigenvalues that are related by the condition $\alpha_\ell + \beta_\ell = \theta$.

Next we deal with the fact that the vectors $\vert \ell \rangle_A$, $\vert \ell \rangle_B$ may not span all of ${\cal H}_A$, ${\cal H}_B$.  To do this define
\bea
\nonumber && {\cal H}_{A_1} = {\rm span} \, \lbrace \vert \ell \rangle_A \rbrace \hspace{7mm} \hbox{\rm (the vector space spanned by $\vert \ell \rangle_A$)} \\
\nonumber && {\cal H}_{A_2} = \left({\cal H}_{A_1}\right)_\perp \hspace{1.3cm} \hbox{\rm (the orthogonal complement of ${\cal H}_{A_1}$ inside ${\cal H}_A$)} \\
&& {\cal H}_{B_1} = {\rm span} \, \lbrace \vert \ell \rangle_B \rbrace  \hspace{7mm} \hbox{\rm (the vector space spanned by $\vert \ell \rangle_B$)} \\
\nonumber && {\cal H}_{B_2} = \left({\cal H}_{B_1}\right)_\perp \hspace{1.3cm} \hbox{\rm (the orthogonal complement of ${\cal H}_{B_1}$ inside ${\cal H}_B$)}
\eea
The vectors $\vert \ell \rangle_A$ form an (overcomplete) basis for ${\cal H}_{A_1}$.  Since $u_A$ preserves this basis, it follows that $u_A$ is block-diagonal
when acting on ${\cal H}_{A_1} \oplus {\cal H}_{A_2}$.
\be
u_A = \left(\begin{array}{c|c}
u_{A_1} & 0 \\
\hline
0 & u_{A_2}
\end{array}\right)
\ee
Likewise $u_B$ preserves ${\cal H}_{B_1}$, so it is block-diagonal when acting on ${\cal H}_{B_1} \oplus {\cal H}_{B_2}$.
\be
u_B = \left(\begin{array}{c|c}
u_{B_1} & 0 \\
\hline
0 & u_{B_2}
\end{array}\right)
\ee
At this point it's convenient to diagonalize $u_{A_1}$.  Suppose $u_{A_1}$ has distinct eigenvalues
\bea
\nonumber
&& e^{i a_1} \quad \hbox{\rm with degeneracy $m_1$} \\
&& e^{i a_2} \quad \hbox{\rm with degeneracy $m_2$} \\
\nonumber
&& \qquad\qquad \vdots \\
\nonumber
&& e^{i a_s} \quad \hbox{\rm with degeneracy $m_s$}
\eea
We likewise diagonalize $u_{B_1}$ with distinct eigenvalues $e^{i b_i}$ and corresponding degeneracies $n_i$, $i = 1,\ldots,s$.
The eigenvalues are related by the condition $a_i + b_i = \theta$.  (This condition implies that $u_{A_1}$ and $u_{B_1}$ have the same
number of distinct eigenvalues.)  In this diagonal basis $u_A$ and $u_B$ have the form
\bea
\label{uAuB}
&& u_A = \left(\begin{array}{ccc|c}
e^{i a_1} \identity_{m_1 \times m_1} &&& \\
& \ddots && \\
&& e^{i a_s} \identity_{m_s \times m_s} & \\
\hline
&&& u_{A_2}
\end{array}\right) \\[8pt]
\nonumber
&& u_B = \left(\begin{array}{ccc|c}
e^{i b_1} \identity_{n_1 \times n_1} &&& \\
& \ddots && \\
&& e^{i b_s} \identity_{n_s \times n_s} & \\
\hline
&&& u_{B_2}
\end{array}\right)
\eea
Now we turn to system $C$.  Let's order the labels $\ell$ so that
\bea
\nonumber
&& \hbox{\rm the first $o_1$ have $u_A \vert \ell \rangle_A = e^{i a_1} \vert \ell \rangle_A$ and $u_B \vert \ell \rangle_B = e^{i b_1} \vert \ell \rangle_B$} \\
&& \hbox{\rm the next $o_2$ have $u_A \vert \ell \rangle_A = e^{i a_2} \vert \ell \rangle_A$ and $u_B \vert \ell \rangle_B = e^{i b_2} \vert \ell \rangle_B$} \\
\nonumber
&& \qquad\quad \vdots \\
\nonumber
&& \hbox{\rm the last $o_s$ have $u_A \vert \ell \rangle_A = e^{i a_s} \vert \ell \rangle_A$ and $u_B \vert \ell \rangle_B = e^{i b_s} \vert \ell \rangle_B$}
\eea
Then we can introduce stabilizers
\bea
&& s_{AC} = e^{i \theta'} \, u_A \otimes \identity_B \otimes \left(\begin{array}{ccc}
e^{-i a_1} \identity_{o_1 \times o_1} && \\
& \ddots & \\
&& e^{-i a_s} \identity_{o_s \times o_s}
\end{array}\right) \\[8pt]
\nonumber
&& s_{BC} = e^{i \theta''} \, \identity_A \otimes u_B \otimes \left(\begin{array}{ccc}
e^{-i b_1} \identity_{o_1 \times o_1} && \\
& \ddots & \\
&& e^{-i b_s} \identity_{o_s \times o_s}
\end{array}\right)
\eea
that by construction satisfy
\be
s_{AC} \vert \psi \rangle = e^{i \theta'} \vert \psi \rangle \qquad\quad s_{BC} \vert \psi \rangle = e^{i \theta''} \vert \psi \rangle
\ee
If we choose the phases to satisfy $\theta' + \theta'' = \theta$, then we also have
\be
\label{CommonCenter}
s_{AC} s_{BC} = s_{AB}
\ee
In other words, for a state obtained by purifying a separable density matrix, an $AB$ stabilizer can always be thought of as a combination of an $AC$ stabilizer with a $BC$ stabilizer.

We can turn this statement about stabilizers into a statement about the two-party pure-state entanglement that is present in the purification.  (This perspective makes an appearance in section 2.7 of \cite{Jiang:2023uyj}.)
Thinking of $s_{AC}$, $s_{BC}$, $s_{AB}$ as representatives of equivalence classes in the pure-state entanglement groups $E_{AC}$, $E_{BC}$, $E_{AB}$, namely
\be
e_{AC} = [s_{AC}] \in E_{AC}, \quad e_{BC} = [s_{BC}] \in E_{BC}, \quad e_{AB} = [s_{AB}] \in E_{AB}
\ee
we have
\be
\label{CommonCenter2}
e_{AC} e_{BC} = e_{AB}
\ee
That is, for a state obtained by purifying a separable density matrix, $AB$ entanglement can always be thought of as a combination of $AC$ entanglement and $BC$ entanglement.

Entanglement satisfying (\ref{CommonCenter2}) is allowed, but it takes a very special and restricted form \cite{Jiang:2023uyj}.  Projecting (\ref{CommonCenter2}) onto system $A$,
we see that $g_A = \pi_A(e_{AC})$ is an element of both $\pi_A(E_{AC})$ and $\pi_A(E_{AB})$.  It was shown in \cite{Jiang:2023uyj} that distinct two-party entanglement groups
must commute element-by-element, which means that $g_A$ must be in the center of both groups.\footnote{The same reasoning leads to (\ref{Z}) in the present paper.}
\be
g_A \in Z\big(\pi_A(E_{AB})\big) \cap Z\big(\pi_A(E_{AC})\big)
\ee
An isomorphism theorem\footnote{See (51) in \cite{Jiang:2023uyj}.  The theorem is analogous to (\ref{isomorphism}) in the present paper.}
implies that elements of the common center act isomorphically on systems $A$, $B$, $C$.

\section{Werner state example\label{appendix:Werner}}
To illustrate the controlled unitary discussed in section (\ref{sect:A(BC)}), we use the example of the Werner state \cite{Werner:1989zz}, defined for two qubits as
\begin{equation}
W = p \, \dyad{\text{Bell state}} + \frac{1-p}{4} \, \identity_{4\times 4}
\end{equation}
where $p \in [0,1]$. But first, let's look at the entanglement group.  The two-party stabilizer group $\widetilde{S}_{AB}$ consists of matrices
\begin{equation}
\widetilde{s}_{AB} = e^{i \theta} \begin{pmatrix}
\alpha & \beta \\
\gamma &\delta
\end{pmatrix}
\otimes
\begin{pmatrix}
\bar{\alpha} & \bar{\beta} \\
\bar{\gamma} & \bar{\delta}
\end{pmatrix} \qquad\quad \begin{pmatrix}
\alpha & \beta \\
\gamma &\delta
\end{pmatrix} \in U(2)
\end{equation}
(an overall phase, times the tensor product of a unitary matrix with its complex conjugate).
To see that this is a stabilizer, note that $\widetilde{s}_{AB} \vert \, \hbox{\rm Bell state} \rangle = e^{i \theta} \vert \, \hbox{\rm Bell state} \rangle$, so the two terms in $W$ are
separately invariant under conjugation by $\widetilde{s}_{AB}$.  This means the two-party stabilizer group is $\widetilde{S}_{AB} = U(1) \times U(2)$.  One-party stabilizers
$\widetilde{s}_A$, $\widetilde{s}_B$ are just phases times the identity operator, so $\widetilde{S}_A = \widetilde{S}_B = U(1)$.  The entanglement group is therefore
\be
\EtildeAB = \big(U(1) \times U(2)\big) / \big(U(1) \times U(1)\big) = U(2) / U(1) = PSU(2)
\ee

The Werner state is separable when $0 \leq p \leq 1/3$ and non-separable for $1/3 < p \leq 1$.
Our goal is to display a block-diagonal unitary matrix (\ref{block}) which completely disentangles system $B$, but
only in the range $0 \leq p \leq 1/3$ where the Werner state is separable.

We begin by presenting the Werner state in the form
\bea
\nonumber
W & = & \frac{p}{2} \, \Big(\dyad{00}{00} + \dyad{11}{11} + \dyad{--} + \dyad{++} + \dyad{\circ\times} + \dyad{\times\circ} \Big) \\
\label{Werner}
& & + (1 - 3p) \, \frac{1}{4} \, \identity_{4\times 4}
\eea
where $+-$ is the $x$-basis and $\times\circ$ is the $y$-basis: $\ket{+}=\big(\ket{0}+\ket{1}\big)/\sqrt{2}$ and $\ket{\times} = \big(\ket{0}+i\ket{1})\big/\sqrt{2}$, etc.  This presents the state as a sum of
tensor products with positive coefficients, except for the the last term where it turns negative precisely when $p > 1/3$.  Assuming the state is separable ($p \leq 1/3$), it can be purified by introducing
an auxiliary 10 dimensional Hilbert space, to
\bea
\nonumber
\ket{\psi} & = & \sqrt{\frac{p}{2}}\Big(\ket{000}+\ket{111}+\ket{--2}+\ket{++3}+\ket{\circ\times4}+\ket{\times\circ5}\Big) \\
& & + \sqrt{\frac{1-3p}{4}}\Big(\ket{006}+\ket{017}+\ket{108}+\ket{119}\Big), \label{werner_sep_pure}
\eea
Note that this purification is not minimal.  It is only a purification of the Werner state if $p \leq 1/3$.  (If $p > 1/3$,
there are imaginary coefficients, and tracing out the auxiliary system will produce a coefficient $\left \vert {1 - 3p \over 4}
\right \vert$ that differs by a sign from (\ref{Werner}).) On the enlarged Hilbert space, we can write down a block-diagonal unitary by inspection:
\bea
\nonumber
&& {\cal U}_{BC} = \identity_A\otimes\Big[\Big(\ketbra{0}{1}+\ketbra{1}{0}\Big)\otimes \dyad{1} + \Big(\ketbra{0}{-}+\ketbra{1}{+}\Big)\otimes\dyad{2} \\
\nonumber
&&  + \Big(\ketbra{0}{+}+\ketbra{1}{-}\Big)\otimes\dyad{3} + \Big(\ketbra{0}{\times}+\ketbra{1}{\circ}\Big)\otimes\dyad{4} + \Big(\ketbra{0}{\circ}+\ketbra{1}{\times}\Big)\otimes\dyad{5} \\
&& +\Big(\ketbra{0}{1}+\ketbra{1}{0}\Big)\otimes\Big(\dyad{7}+\dyad{9}\Big)\Big] + \identity_A \otimes \identity_B \otimes \Big(\dyad{0} + \dyad{6} + \dyad{8}\Big)
\eea
Applying this unitary takes (\ref{werner_sep_pure}) to the claimed tensor product form.
\begin{equation}
\label{werner_factor}
{\cal U}_{BC}\ket{\psi} = \ket{0}_B\otimes\Big[ \sqrt{\frac{p}{2}}\Big(\ket{00}+\ket{11}+\ket{-2}+\ket{+3}+\ket{\circ4}+\ket{\times5}\Big) + \sqrt{\frac{1-3p}{4}}\Big(\ket{06}+\ket{07}+\ket{18}+\ket{19}\Big) \Big]_{AC}
\end{equation}
Of course the Werner state can be purified for any value of $p$.  Indeed a minimal purification, valid for any $0 \leq p \leq 1$, is
\be
\label{w_min_pure}
\ket{\psi} = \sqrt{\frac{1-p}{4}} \, \left[ {1 \over \sqrt{2}} \big( \ket{00} - \ket{11} \big) \otimes \ket{0}
+ \ket{01} \otimes \ket{1} + \ket{10} \otimes \ket{2} \right]
+ \sqrt{\frac{1+3p}{8}} \, \big( \ket{00} + \ket{11} \big) \otimes \ket{3}
\ee
If $p \leq 1/3$ this can be related to (\ref{werner_sep_pure}) by a unitary on system $C$ and can be mapped to
(\ref{werner_factor}) by ${\cal U}_{BC}$.  However if $p > 1/3$ the Werner state is not separable and no such unitary
on $C$ exists.

\section{Entanglement groups for ${\cal U}_{BC} \vert \psi \rangle$ \label{appendix:UBC}}
As shown in section \ref{sect:A(BC)}, for a separable state, there is a controlled unitary ${\cal U}_{BC}$ that completely disentangles system $B$.  This means we can define a new state
\be
\vert \psi' \rangle = {\cal U}_{BC} \vert \psi \rangle
\ee
that has the form given in (\ref{factorized}).
\be
\vert \psi' \rangle = \vert \chi \rangle_B \otimes \Big(\sum_{\ell = 1}^L \sqrt{p_\ell} \, \vert \ell \rangle_A \otimes \vert \ell \rangle_C\Big)
\ee
The entanglement groups for $\vert \psi' \rangle$ have the following properties.
\begin{enumerate}
\item
Since $\vert \psi \rangle$ and $\vert \psi' \rangle$ are related by a (block diagonal) unitary transformation on the combined $BC$ system, the $A$ -- $BC$ entanglement groups are unchanged.
\be
E_{A(BC)}^{\vert \psi \rangle} = E_{A(BC)}^{\vert \psi' \rangle}
\ee
\item
The $A$ -- $BC$ entanglement group for $\vert \psi' \rangle$ acts trivially on system $B$.
\be
\label{prop2}
E_{A(BC)}^{\vert \psi' \rangle} = E_{AC}^{\vert \psi' \rangle}
\ee
To see this, note that the Schmidt decomposition of the state $\vert \psi' \rangle$ has the form\footnote{This is the Schmidt decomposition for a division into systems $A$
and $BC$.  It's the same as the Schmidt decomposition one would obtain by projecting the state $\vert \psi' \rangle$ into ${\cal H}_A \otimes {\cal H}_C$.}
\be
\vert \psi' \rangle = \sum_{i = 1}^r \sqrt{p_r} \, \vert i \rangle_A \otimes \vert \chi \rangle_B \otimes \vert i \rangle_C
\ee
where $r$ is the rank of $\rho_A = {\rm Tr}_B \rho_{AB}$.  For the purposes of $A$ -- $BC$ entanglement, this means we can restrict our attention to stabilizers of the form
\be
s_{AC}^{\vert \psi' \rangle} = u_A \otimes \identity_B \otimes u_C
\ee
It follows that $E_{A(BC)}^{\vert \psi' \rangle} = E_{AC}^{\vert \psi' \rangle}$.  For the original state $\vert \psi \rangle$ note that the relevant stabilizers
$s_{A(BC)}^{\vert \psi \rangle}$ do act on system $B$, but in a way that is completely fixed by
\be
s_{A(BC)}^{\vert \psi \rangle} = {\cal U}_{BC}^\dagger s_{AC}^{\vert \psi' \rangle} {\cal U}_{BC}
\ee
\item
Since $\vert \psi' \rangle$ is a tensor product, there is no entanglement between $B$ and the combined $AC$ system.
\be
\label{prop3}
E_{B(AC)}^{\vert \psi' \rangle} = \lbrace 1 \rbrace
\ee
\end{enumerate}
These features of the entanglement groups for $\vert \psi' \rangle$ are consequences of having a separable density matrix.

Given (\ref{prop3}), the argument is reversible.  Consider a state $\vert \psi' \rangle$, which without loss of generality we write as\footnote{Any state can be expanded in
a basis of tensor product states, $\vert \psi' \rangle = \sum c_{i\ell} \vert i \rangle_{AB} \otimes \vert \ell \rangle_C$.  Collect the coefficient of $\vert \ell \rangle_C$ and write
it as a unit vector $\vert \ell \rangle_{AB}$ times a magnitude $\sqrt{p_\ell}$.}
\be
\vert \psi' \rangle = \sum_\ell \sqrt{p_\ell} \, \vert \ell \rangle_{AB} \otimes \ell_C
\ee
with ${}_C \langle \ell \vert \ell' \rangle_C = \delta_{\ell \ell'}$.
If $E_{B(AC)}^{\vert \psi' \rangle} = \lbrace 1 \rbrace$, then $\vert \psi' \rangle$ must be a tensor product of the form (\ref{factorized}).  Then, by applying
the inverse of the block-diagonal unitary transformation (\ref{block}) we obtain (\ref{pure2}), which is the general purification of a separable density matrix.

\providecommand{\href}[2]{#2}\begingroup\raggedright\endgroup

\end{document}